\documentclass[pra,twocolumn,twoside,aps]{revtex4-1}

\usepackage{amsmath, amssymb}
\usepackage{bm}
\usepackage{natbib}
\usepackage[dvipsnames]{xcolor}
\usepackage[normalem]{ulem}

\usepackage{graphicx}
\usepackage{float}

\begin{document}
\title{Computational time-of-flight diffuse optical tomography.}


\author{Ashley Lyons$^{1}$, Francesco Tonolini$^2$,  Alessandro Boccolini$^{1}$, Audrey Repetti$^3$,   Robert Henderson$^4$, Yves Wiaux$^3$, Daniele Faccio$^1$}

\affiliation{$^1$School of Physics and Astronomy, University of Glasgow, Glasgow, G12 8QQ, UK}
\affiliation{$^2$School of Computing Science, University of Glasgow, Glasgow, G12 8QQ, UK}
\affiliation{$^3$Institute of Sensors, Signals and System, Heriot-Watt University, Edinburgh EH14 4AS, UK}
\affiliation{$^4$Institute for Micro and Nano Systems, University of Edinburgh , EH8 9YL, Edinburgh, UK}

\begin{abstract}

Imaging through a strongly diffusive medium remains an outstanding challenge in particular in association with applications in biological and medical imaging. Here we propose a method based on a single-photon time-of-flight camera that allows, in combination with computational processing of the spatial and full temporal photon distribution data, to image an object embedded inside a strongly diffusive medium over more than 80 transport mean free paths. The technique is contactless and requires one second acquisition times thus allowing Hz frame rate imaging. The imaging depth corresponds to several cm of human tissue and allows one to perform deep-body imaging, here demonstrated as a proof-of-principle.

\end{abstract}

\maketitle

\begin{figure}[t!]
	\centering
	\includegraphics[width=8cm]{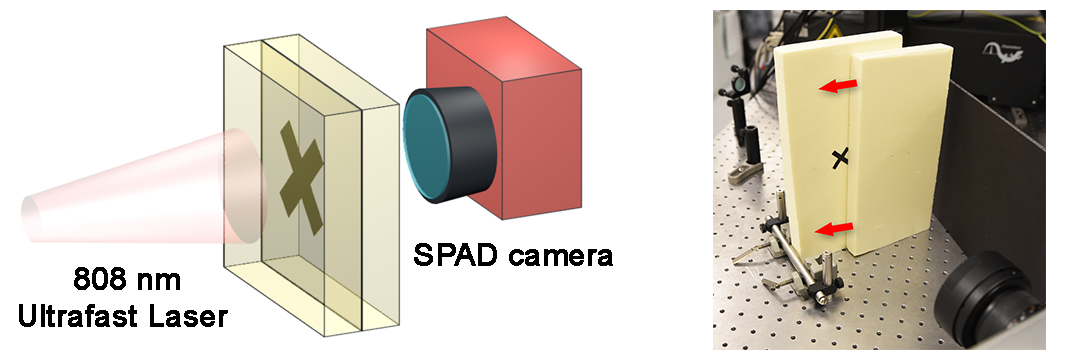}
	\caption{Layout (left) and photograph (right) of the experimental layout. The input laser beam is defocused to a diameter of $\sim5$ cm and is centred on the embedded target (shapes cut out of black  tape). The SPAD camera (visible also in the bottom right hand corner of the photograph) is placed on the opposite side of the diffusive slabs in order to collect the transmitted laser light.}
	\label{fig:layout}
\end{figure}

\begin{figure*}[t!]
	\begin{center}
		\includegraphics[width=1.4\columnwidth]{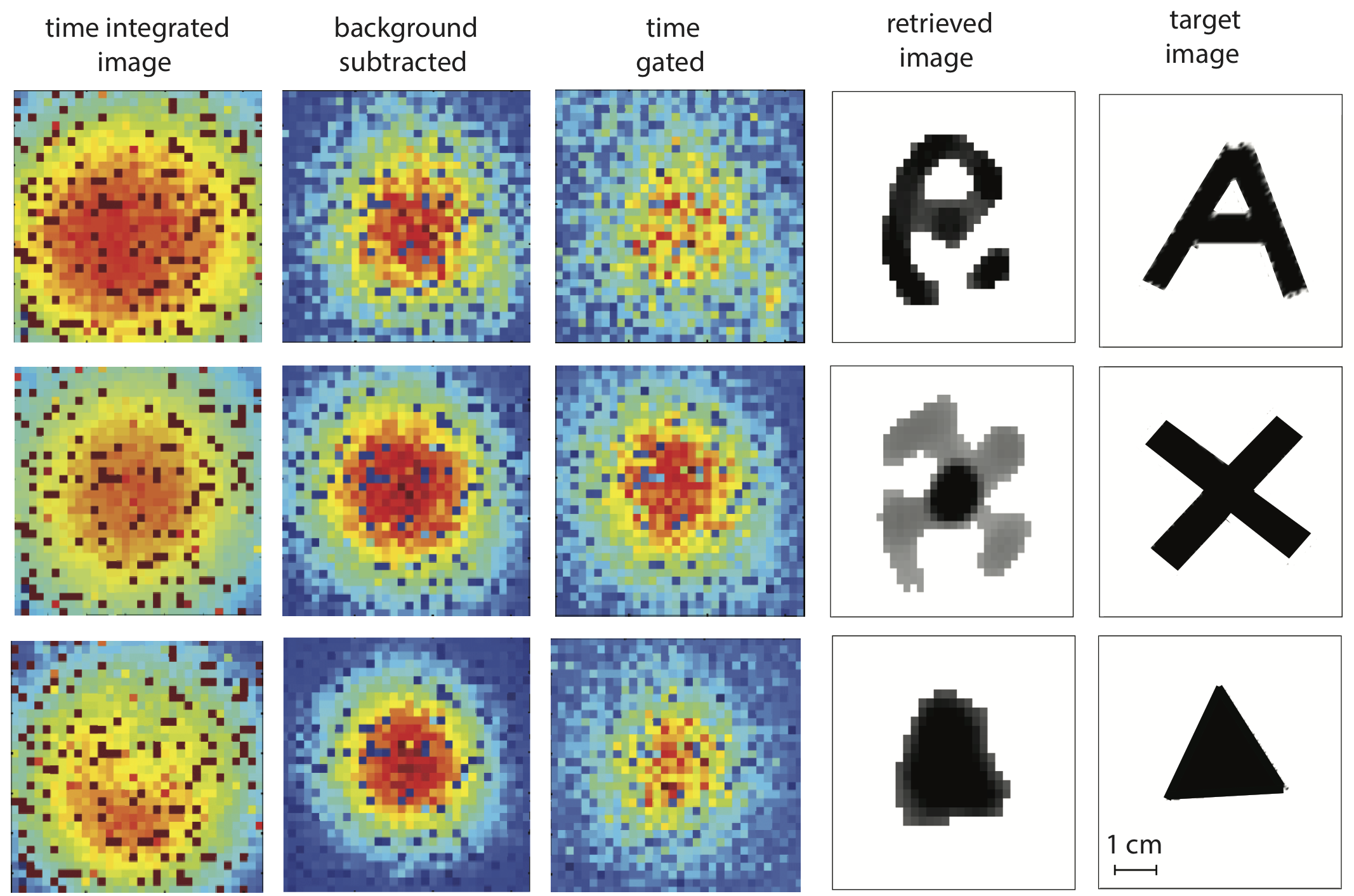}
		\caption{\label{fig:reconstruction}{Main experimental results. The first column shows the object hidden inside the scattering medium. The second and third columns show the recorded image of the transmitted laser beam and time-gated on the first 10 time-bins that are above the noise level respectively. No discernible image is visible, showing that time-gating in an attempt to isolate ballistic photons is not successful. The fourth column shows the retrieved object images using the technique described in the main text. The fifth column shows the original unknown objects (Dark pixels correspond either to dead pixels or to so-called `screamers', i.e. defected pixels with abnormally high dark counts). %
		}}
	\end{center}
\end{figure*}

Visible or near-infrared (NIR) light propagating in turbid media, for example biological tissue or a foggy environment, follows a complicated random path due to multiple scattering. As a consequence, the optical wavefront is severely modified and its intensity is rapidly attenuated in propagation. This leads to the inability of an imaging system to detect an object that is located within and thus obscured by the medium.\\
Recent efforts have also been directed at imaging objects that are located behind or embedded in a scattering medium \cite{behind0,behind1,behind2}. \\
Generally speaking, photons propagating in  a scattering medium can be divided into  \textit{ballistic}, \textit{snake} and \textit{diffusive photons} \cite{wang1991ballistic}. Ballistic and snake photons propagate with no or very little interaction with the scatterers along the direction of the beam. They therefore retain their original  coherence and most of the image information. However, they are also exponentially suppressed and do not survive beyond distances of several cm in biological or highly scattering tissue. A medium of thickness $L$ is considered to be highly diffusive when the transport mean free path $\ell^{*}=1/\mu'_s \ll L$, where $\mu_a$ and $\mu'_s$ are the absorption and reduced scattering coefficients, respectively \cite{Konecky2008,Durduran2010,diffuse-review,shi2017_book} with typical values for biological tissue that are of order $\mu_a\sim 0.05$ 1/cm and $\mu'_s\sim10$ 1/cm ($\ell^{*}\sim 0.1$ cm) \cite{tissue-review}. The transport mean free path represents the distance over which all information on the photon's initial propagation direction is lost.  Measurements of light transmitted through such a material therefore carry very little or no direct image information. Here we focus attention on this propagation regime.\\ 
{The first generation of experiments and methods for diffuse imaging were developed in the late 1980's, early 1990's, establishing the boundaries in terms of maximum imaging depth and resolution \cite{delpy88,wilson89,jacques89,hebden92,delpy95}. Successive generations were aimed at medical tests in a variety of conditions and also in vivo \cite{gibson-review,berg1993medical,gros99,boas2001imaging,torr14,egg14}. The aim of most studies in recent years has  been towards increasing image contrast, depth sensitivity and decreasing acquisition time \cite{diffuse-review,mora14,charbon14,hebden-review}.\\
In the strongly diffusive regime, light will propagate in the form of Photon Density Waves (PDWs) that exhibit many features typical of standard propagating waves, including interference, diffraction and also imaging properties. Imaging properties are essentially determined by the wave-vector associated to PDWs, $\kappa_d=\sqrt{3\mu_a/\ell^{*}}$ \cite{Konecky2008}. For typical biological tissue, $\kappa_d\sim1$ 1/cm, thus limiting imaging resolution to transverse dimensions that are of the same order of magnitude of the medium thickness, e.g. spatial resolutions of the order of 5 cm are achieved in  5 cm thick samples \cite{Ripoll1999,Durduran2010}. This can be improved upon by using computational techniques, e.g. inverse retrieval algorithms \cite{Konecky2008}, or by post-selecting data in the temporal domain \cite{Azizi2009} to achieve resolutions of $\sim$1 cm with realistic scattering parameters. We note that in the latter case, the majority of the temporal information was discarded to filter out data only at one specific temporal slice where the spatial resolution was found to be highest.}\\
%
%
Computational based time-resolved measurements combined with ultrafast imaging have demonstrated to be a promising technique in retrieving information lost in a highly scattering medium, see e.g. \cite{Durduran2010} for a review. Among these, an approach was introduced that builds upon all of the temporally resolved data, named All Photons Imaging (API). API utilizes both spatial and temporal (photon arrival time) components of scattered light and has successfully demonstrated to improve the spatial resolution of an object hidden behind a turbid medium \cite{satat2016all}. \\
We underline that  in all the methods outlined above,  the acquired data (images at the output plane of the scattering medium) shows  a clear shadow that is cast by the hidden object and is always clearly visible, even in the time-integrated image. The effect of the computational methods including also API, is to significantly improve the spatial resolution of the acquired image. Pioneering work was performed by Cai et al. in which the position of 5 mm sized objects embedded within 60 mm of a diffusive medium (2.5 mm transport mean free path) was determined using fiber source/detector pairs and a streak camera \cite{Cai1996}.\\
%
Here we introduce a Time-of-Flight Diffusive Optical Tomography (ToF-DOT) approach to address the problem of imaging an object deeply embedded inside a highly scattering medium. {Unlike the aforementioned methods, we investigate a regime where any signatures, even of the presence of an occluding object within the scattering medium, are too weak to be identified from visual inspection of the raw data alone.} We use the full spatial and temporal information of the photon time-of-flight at each pixel spatial position recorded on a single-photon array detector in combination with a computational retrieval method  in order to estimate the hidden object shape and position. 
The ToF-DOT approach allows to recover the shape of a 2D opaque object hidden inside more than 80 transport mean-free-path lengths of diffusive material (corresponding to several cm of human tissue). We reconstruct mm-sized features  and find that both the spatial and temporal resolution of the camera are key to achieving these results.  Moreover, the technique is sufficiently sensitive to allow data acquisition on time scales of the order of 1 second.\\
\noindent {\bf{Experimental setup.}}
We aim to reconstruct the shape and location of a two dimensional object embedded within a diffusive medium by performing spatially and temporally resolved intensity measurements of femtosecond light pulses transmitted through the medium. Our experimental setup is illustrated in Fig. \ref{fig:layout}.
We use a pulsed laser source with a wavelength of 808 nm, 120 fs temporal pulse duration, 80 MHz repetition rate and 1 W average power defocused to 2.5 cm radius spot size, thus corresponding to an illumination fluence of 0.5 mW/mm$^2$. This  illuminates an object inside a medium consisting of two slabs of polyurethane foam, each 2.5 cm thick, with absorption and reduced scattering coefficients at the illumination wavelength measured from a single point, time-resolved measurement to be $\mu_a$ = 0.09 cm$^{-1}$ and $\mu'_s$ = 16.5 cm$^{-1}$ (see Methods for details). The material thus has a transport mean free path $\ell^{*}=600$ $\mu$m that is nearly two orders of magnitude smaller than the total thickness of the material, $L=5$ cm.\\
The laser pulses are transmitted through the diffusive medium and some of the light is absorbed by a hidden object placed between the two slabs. Black  tape was used to create hidden targets of different shapes, e.g. letters (A, X), triangles or double lines. The transmitted light is collected by a camera composed of a $32 \times 32$ array of Single Photon Avalanche Diode (SPAD) detectors (commercialised by Photon Force Ltd.), each one operating in Time Correlated Single Photon Counting (TCSPC) mode with 55 ps resolution \cite{gariepy2015single}. The SPAD camera therefore collects three-dimensional data: two spatial dimensions with {$N \times N = 32 \times 32$} pixel resolution and one temporal dimension ({$T=230 \times 55$} ps time-bins).\\
The first column in Fig.~\ref{fig:reconstruction} shows typical examples of time-integrated transmission images measured with the camera for various objects with feature sizes of order $\sim1$ cm (shown in the last column). We also show time-gated images in the third column, attempting to isolate any eventual ballistic photons. The notable feature of these images is that in none of these is it possible to visually determine the presence (or absence) of an object embedded inside the medium. \\
{\bf{Computational retrieval model.}}
As noted above, the material has a transport mean free path $\ell^{*}=600$ $\mu$m that is two orders of magnitude smaller than the total thickness of the material. This places light in the strongly diffusive regime, which in turns allows us to model the photon propagation inside the diffusive medium using a diffusion approximation \cite{yoo1990does}. Light within the medium essentially behaves like heat, following the steepest descent of the scalar gradient weighted by the diffusivity, with an additional loss effect due to photon absorption \cite{wang2007diffuse}. The diffusion equation in the context of photon diffusion is expressed as 
\begin{equation}
	\label{eq1}
	c^{-1}\frac{\partial \Phi(\vec{r},t)}{\partial t}+\mu_{a}\Phi(\vec{r},t)- D\nabla \cdot \left[\nabla\Phi(\vec{r},t)\right]=S(\vec{r},t)
\end{equation}
where $c$ is the speed of light in the medium, $\vec{r}$ is the spatial position, $t$ is the temporal coordinate, $\Phi(\vec{r},t)$ is the photons flux, $S(\vec{r},t)$ is a photon source and $D$ is a term which includes the absorption coefficient $\mu_{a}$ and the reduced scattering coefficient $\mu'_{s}$ and in this work does not depend on $\vec{r}$ or $t$: $ D = \big( 3 (\mu_a + \mu'_s) \big)^{-1} $. A full derivation of Eq.~(\ref{eq1}) beginning with the radiative transfer equation can be found in \cite{wang2007diffuse}.
%
%
For the case of a highly localized (in space and time) input laser pulse, Eq.~(\ref{eq1}) has an analytical solution given by \cite{wilson89}
\begin{eqnarray}
	\label{eq3}
	&\Phi(\vec{r},t;\vec{r}',t')&= \frac{c}{\left[4\pi Dc(t-t')\right]^{3/2}} \times \nonumber \\ 
	&&\exp\left[-\frac{\lvert\vec{r}-\vec{r}'\rvert ^{2}}{4Dc(t-t')}\right]\exp\left[-\mu_{a}c(t-t')\right].
\end{eqnarray}
Here, $\vec{r}'$ and $t'$ identify the position and time of the input laser pulse. Eq.~\eqref{eq3} describes the evolution of a delta function in time and can be applied to an extended light source (see Methods). \\ 
The image-retrieval model can be described as an inverse problem, where the aim is to estimate the shape of the hidden object $x \in \mathbb{R}^{N \times N}$, from the 3D  (2 spatial and 1 temporal dimension) observation obtained by the SPAD camera, denoted by $Y$. We have $Y = \mathcal{A}(x)$, where $\mathcal{A}$ is the linear operator mapping the original 2D image $x$ to the 3D measurements.\\ 
The first step is to compute a forward model using Eq.~(\ref{eq3}) to simulate light propagation from the input plane to the object plane and then, after masking with a guess estimate (for example, at the first step this can be a simple flat, zero-amplitude distribution), propagation from the object plane to the diffuse medium output. This numerical solution is then compared to the actual measurement by evaluating a cost function. This function is in turn used to modify the shape of the object guess function and is minimised through an iterative process of solving the forward model with the adapted guess target function. Full details of the forward model and iterative cost function minimisation are given in the Methods section.\\
\noindent {\bf{Results and Discussion.}}
The absorption and scattering parameters of the polyurethane (PU) were measured before the experiment by fitting the temporal diffusion to Eq.~\eqref{eq3} (see Methods). We then placed objects of various shapes and out of black tape at the interface between the two slabs of material (as shown in Fig.~\ref{fig:layout}). \\
\begin{figure}[b]
	\centering
	\includegraphics[width=8cm]{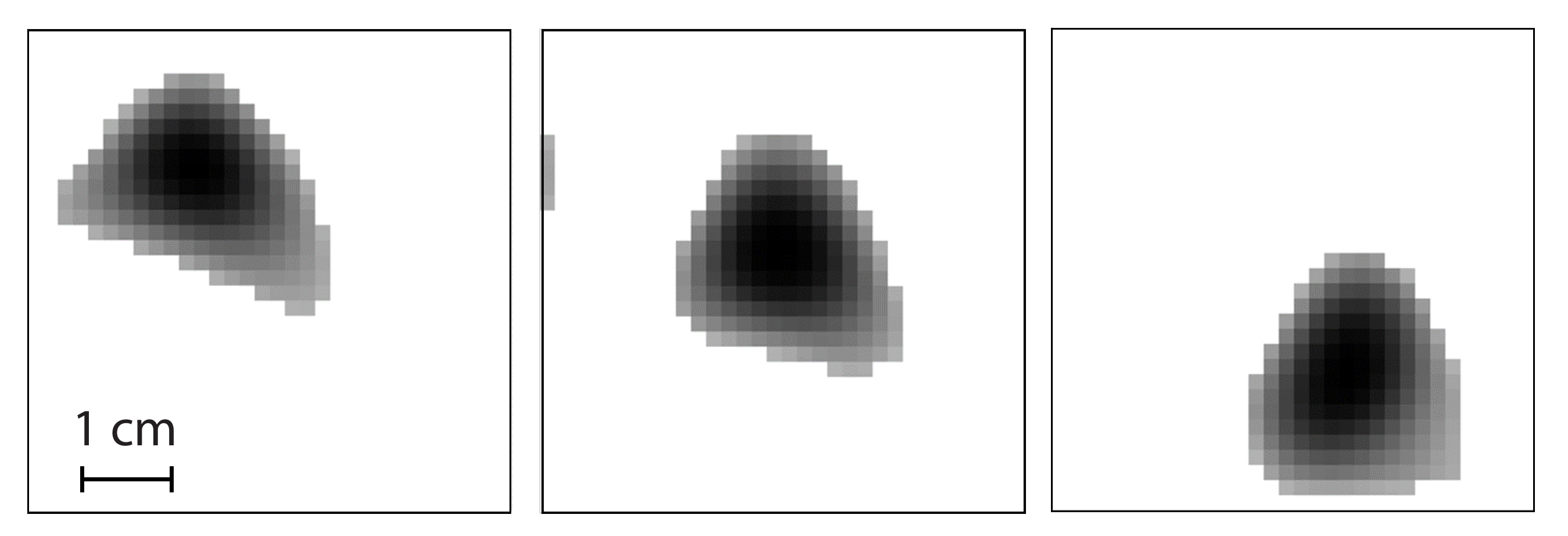}
	\caption{Tracking of a hidden object positioned at different positions inside the diffusive medium and ``captured'' at 1.5 s intervals.}
	\label{fig:tracking}
\end{figure} 
Different shapes were tested: The letters ``A'' and ``X"  and a triangle shape.  Fig.~\ref{fig:reconstruction} shows the results with data for the three different objects on three different rows. As indicated in the figure, the first columns show the raw data, as recorded directly on the SPAD camera (imaging the output side of the diffusive material), the second column shows the background subtracted data (i.e. the data after subtracting out a measurement taken with the laser off), and the third column shows a time-gated image (taken by isolating the first 10 temporal bins of data that rise above the noise floor). As can be seen, there is no discernible information in these time-gated photons and the images actually resemble very closely the total time-integrated images. Reducing the number of time-bins selected to perform the gating leads only to a reduction of the overall signal, with no further information on the presence or shape of the occluded object. Finally, the last two columns show the retrieved image of the occluded object and the actual ground truth for the object. It can be noticed that the ToF-DOT allows to correctly assess the presence of the occluded object and also provides a good qualitative agreement with the actual object shape. Whilst the time integrated or time-gated camera recordings do not show any distinct shadows and thus do not allow to guess the shape or position of the hidden objects, the method is also sensitive to the exact position of the object as shown in Fig.~\ref{fig:tracking} with an example where the ``triangle'' is shifted in three different positions whilst everything else (i.e. laser illumination and camera position) remains unchanged. The data for these images was also acquired at 1.5 second intervals, showing the potential for tracking of changes within the medium in real-time. We do also note however, that the retrieval algorithm used to estimate the object shape of interest was performed offline. It required a few minutes to converge for each image on a standard laptop computer, using a Matlab implementation. This, could be reduced to sub-second timescales by employing parallel computing methods.\\
\begin{figure}[t]
\centering
\includegraphics[width = 0.5\textwidth]{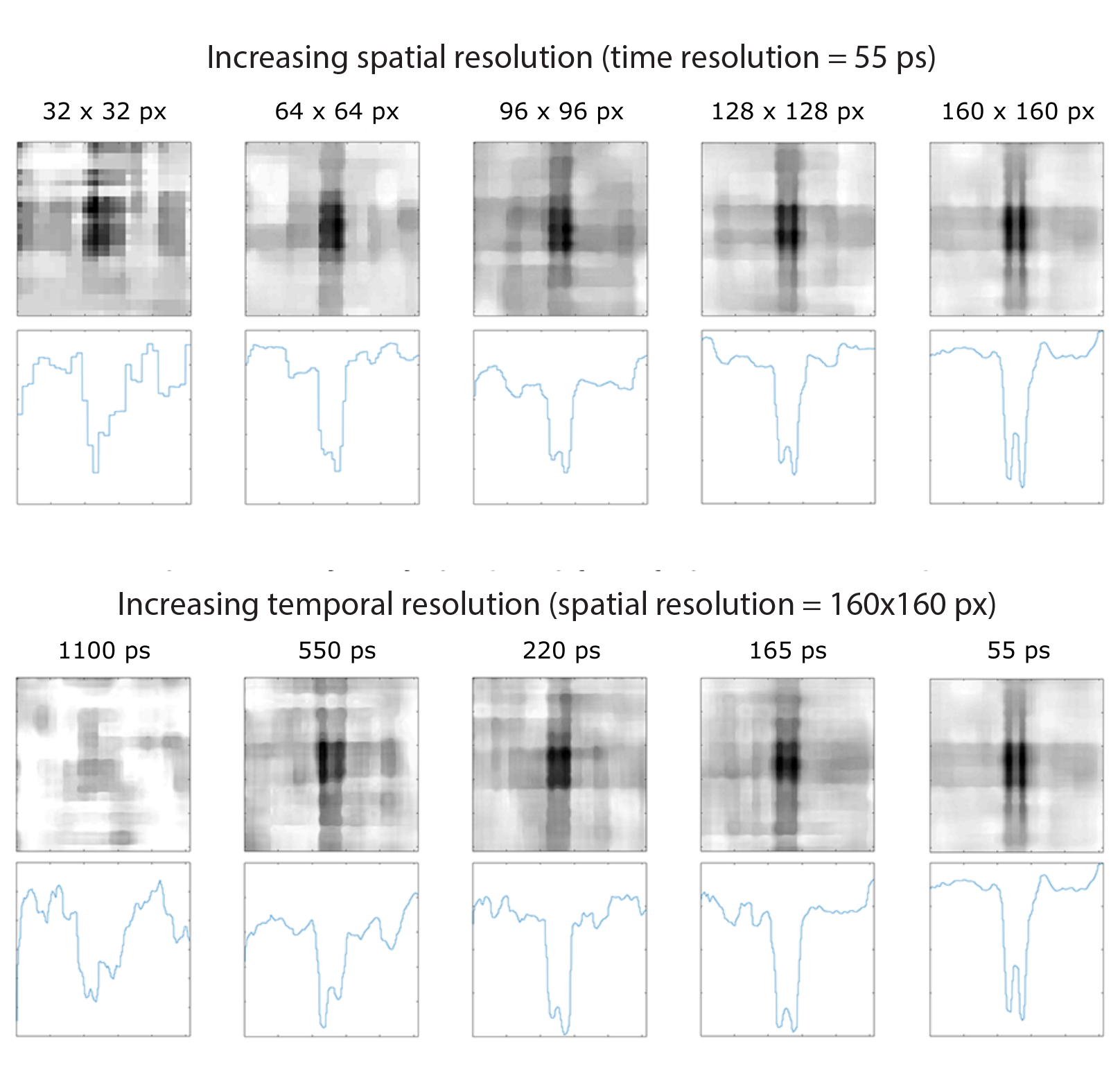}
\caption{Numerical simulations of the reconstruction of a hidden object (0.5 mm thickness, 5 mm height, separated by 1 mm). The top row shows the impact of increasing spatial resolution of the camera for a fixed temporal resolution (55 ps). Pixel densities are shown above each figure illustrating the reconstructed object and a vertical binning of the image to highlight the spatial resolution along the horizontal direction. The bottom row shows the impact of increasing the temporal resolution of the camera for a fixed spatial resolution (160$\times$160 pixels). The images have not been thresholded (see Methods) in order to highlight the effects of spatial and temporal resolution}.
\label{fig:res_sim}
\end{figure}
\begin{figure}[t]
\centering
\includegraphics[width = 8.5cm]{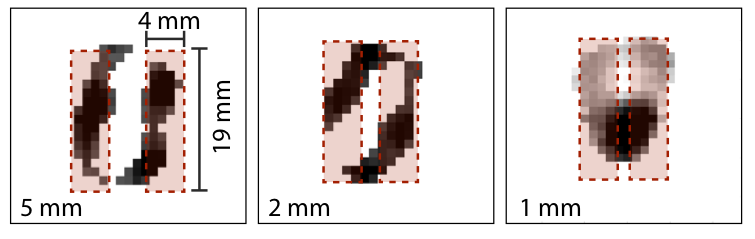}
\caption{Experimental results: Imaging double-lines with 32x32 pixel density: the figures show measurements for three different separations (5, 2, 1 mm as indicated in the figures) between the two vertical lines (geometry indicated in the leftmost figure). The shaded areas show the ground truth positions of the vertical lines. }
\label{fig:lines}
\end{figure}
We underline once more that in the operating conditions used here, typical e.g. of a thick sample of human tissue (that has an average scattering length of 15-20 cm$^{-1}$ \cite{Jacques2013}), standard methods such as time gating of ballistic or snake photons \cite{wang1991ballistic} do not allow to recognise the shape of any object considered in our experiment. Indeed, the 5 cm thick material of our experiment is equivalent to 83 transport mean free paths and this effectively eliminates all ballistic photons. \\
As can be seen in these results, in all cases the algorithm struggles to reconstruct features such as sharp edges, with a resolution that is limited here to $\sim0.5$ cm. However, we performed a series of numerical simulations based on using the forward model with a hidden object that is composed of two vertical stripes, 0.5 mm thick, 5 mm long and separated by 1 mm. These stripes were embedded in a diffusive medium that has the same $\mu'_s$ and $\mu_a$ as in the experiment at a distance of 2.5 cm from the output surface. We then add noise to the output image in order to simulate the camera noise and then use the data in the ToF-DOF reconstruction algorithm to test the ability to correctly identify the 1 mm gap between two stripes. These tests were performed with increasing spatial (first row in Fig.~\ref{fig:res_sim} with temporal resolution fixed at 55 ps) and temporal resolution (first row in Fig.~\ref{fig:res_sim} with spatial resolution fixed at 160$\times$160 pixels). We can see that increasing spatial or temporal resolution on the camera leads to an increase in the resolution of the final retrieved image of the occluded object. In particular, with 55 ps time bins and 160$\times$160 pixels, feature sizes as small as 1 mm are clearly visible. \\
On the basis of this finding we performed an experiment aimed at testing our current spatial resolution capability: two vertical stripes (4 mm thickness, 19 mm height) were used, separated by 5, 2 and 1 mm with a fixed pixel count of 32$\times$32. The results are shown in Fig.~\ref{fig:lines} that displays the 2D reconstructions. A good qualitative agreement with the ground truth (shaded rectangles) is observed for all separations although at 1 mm separation, clear artefacts start to appear, e.g. the two slits are fused together in the lower half. Nevertheless, the retrieval is still able to correctly recognise the existence and overall shape/position of the slits. These measurements therefore highlight current limitations of our approach (difficulty with highly asymmetric features that have details in the sub-mm region) but also the potential to resolve close to mm features in the occluded object.\\
We note that the 55 ps temporal resolution of the camera corresponds to 1.5 cm in free space. But in diffusive propagation one should consider the PDW as the wave propagating information through the system and this travels at a much slower speed, requiring several nanoseconds to transit 5 cm, corresponding to a camera resolution  $\sim0.5$ mm. This simplified reasoning seems to agree with our findings that resolution is limited at the mm scale. 
In more detail, if we consider the case in which there is no embedded object in the medium, then each pixel on the camera will record a temporal profile for the photon arrival times at the output that is exactly described by Eq.~\eqref{eq3}. Moreover, we know that early arrival times correspond to photons taking shorter, i.e. more direct paths to the camera and longer times correspond to photons that travel longer distances due to multiple scattering effects. In the presence of an absorbing object with a spatially extended shape,  photon paths that intersect the object will be blocked and will therefore be absent from the final temporal measurement. This in turn will lead to temporal profiles that deviate slightly from Eq.~\eqref{eq3}. Furthermore, each spatial pixel on the camera is collecting a different subset of photon paths from the medium so that in general, the deviations from the perfect temporal profile Eq.~\eqref{eq3} will vary from pixel to pixel. The shape of the object is therefore encoded in this spatially varying temporal information. One can no longer resort to an exact analytical relation to describe the temporal profile modifications at each pixel. These are however, still fully determined from the solution of the diffusion equation Eq.~\eqref{eq3} when including the embedded, absorbing object. The retrieval algorithm is therefore iteratively reconstructing the shape of the object that best matches the equation predictions to the measured temporal modifications of the photon signal at each pixel.

\noindent {\bf{Conclusions.}}
We have proposed a computational imaging technique for detecting hidden objects that are completely immersed in a highly scattering medium. The method relies on the full, spatially resolved,  time-of-flight information of the photons that are transmitted through the medium and recorded with a photon-counting SPAD camera. The high sensitivity of the camera allows fast acquisition times on the order of 1 second and precise ToF timing. We have shown that by introducing the full ToF information, we can resolve features in the 1-5 mm range and that this can be improved by increasing both spatial \emph{and} temporal resolution of the camera. High spatial and temporal resolution SPAD cameras are currently being developed that also have improved pixel fill factors \cite{henderson2018} ($\sim 60\%$ compared to the $\sim 1\%$ used in these experiments) and thus promise even shorter acquisition times and higher resolutions. We note that Intensified CCD cameras are also available with the required 100-200 ps temporal resolution and could be used to perform similar measurements to those shown here.\\
This work was carried out under the assumption that the medium is homogeneous, which will not in general be true in the case of actual biological tissue or organs. Future work will therefore need to consider the impact of this for example by including a detailed model of the inhomogeneity in the forward model or by searching for methods for adapting the inverse retrieval. \\

\noindent {\bf{Acknowledgements.}}\\
D.F. acknowledges financial support the Engineering and Physical Sciences Research
Council (EPSRC, UK, Grants No. EP/M006514/1 and No. EP/M01326X/1).\\

\noindent {\bf{Data and code availability.}} All codes and data used in this work are available from DOI: http://dx.doi.org/10.5525/gla.researchdata.642 and from Github: {https$://$github.com/ftonolini45/\\
Computational{\_}ToF{\_}Diffuse{\_}Optical{\_}Tomography.}\\

\noindent {\bf{Methods.}}\\
 {\bf{Experiment details:}}
A femtosecond laser source delivers 130 fs pulses at 808 nm with a repetition rate of 80 MHz and 1 W average power. A small fraction is reflected off a beam splitter to an optical constant fraction (OCF) discriminator, while most of the energy is directed towards the scattering medium after the beam has been expanded using a diverging lens. On the other side of the sample the SPAD camera is collecting the transmitted light having interacted with both the scattering medium and the hidden object placed inside. 

The SPAD camera is composed of 32$\times$32 array of SPAD detectors (100 $\mu$m pitch, 8 $\mu$m pixel active area diameter) each one operating in TCSPC mode. Each individual SPAD can detect the time of arrival of a single photon with a time resolution of about 55 ps and impulse response function (IRF) of 120 ps. The OCF output provides the trigger signal for the SPAD camera. The transmitted light is imaged to the SPAD array through an 8 mm focal length photographic lens (Samyang 8mm f/3.5 UMC Fish-eye), and the camera is kept at a fixed distance such that the correspondent field of view (FOV) is covering an area larger than the hidden object dimensions. \\

\begin{figure}[h!]
	\begin{center}
		\includegraphics[width=0.9\columnwidth]{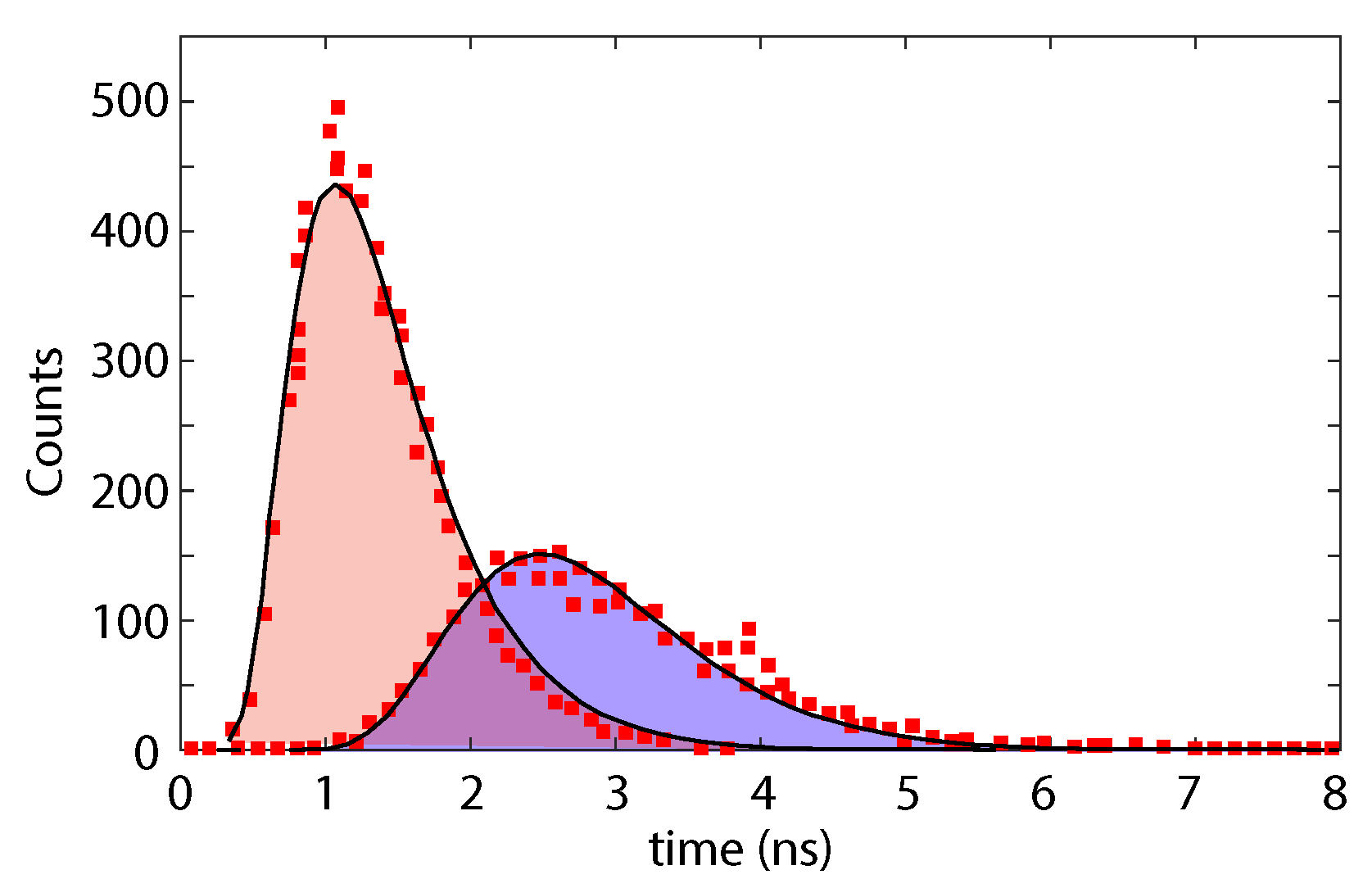}
		\caption{\label{fig:scattering_coefficient}
			{Single pixel temporal histograms of the photon arrivals transmitted through 2.5 cm and 5 cm of material. The thicker solid lines show the best fits with Eq.~\eqref{eq3}, with $\mu'_s=16.5$ 1/cm used as a fitting parameter.
		}}
	\end{center}
\end{figure}

\noindent {\bf{Characterisation of the scattering medium:}} The reduced scattering coefficient $\mu'_s$ of the PU foam was estimated using the experimental setup shown in Fig.~\ref{fig:layout}, where the sample consisted of two polyurethane slabs (without any hidden objects). The absorption coefficient $\mu_a$ was measured prior to the experiment using a spectrophotometer but can be verified also from the same time-resolved measurement used to estimate $\mu'_s$. The reduced scattering coefficient is estimated by comparing the measured temporal broadening and optical delay of a laser pulse with the expected model given by the diffusion approximation, Eq.~\eqref{eq3}. Fig.~\ref{fig:scattering_coefficient} shows the raw data measured at a single pixel for a medium of total thickness 5.0 cm and 2.5 cm: in both cases the  coefficients that best fit the experimental data were found to be $\mu_a=0.09$ 1/cm and $\mu'_s=16.5$ 1/cm.\\
 \noindent {\bf{Retrieval algorithm:}} 
As described in the Computational retrieval model section, we denote with $Y \in \mathbb{R}^{N \times N \times T}$ the 3D measurements, containing $T$ images of size $N\times N$, and described by the following forward model:
\begin{equation}\label{pb_inv}
Y = \mathcal{A}(x) + W.
\end{equation}
In Eq.~\eqref{pb_inv}, $W$ is a realization of an additive random noise 
 and 
$\mathcal{A} \colon \mathbb{R}^{N \times N} \to \mathbb{R}^{N \times N \times T}$ is the observation operator mapping linearly the hidden object to the 3D measurements.
This operator models the acquisition process described in the Computational retrieval model section.
The inverse problem defined by Eq.~\eqref{pb_inv} is ill-posed and requires the development of adapted tools in order to be solved.  
During the last decades, optimization techniques have been developed to tackle such  problems arising in different signal processing fields \cite{combettes2011proximal, komodakis2014_playing_duality}. In this context, the unknown object is defined 
as a minimizer of an objective function made of a sum of two terms: the data fidelity term related to the forward model and the regularization term incorporating \emph{a priori} information we have on the target object (e.g. piece-wise constant image). 
Therefore, we proceed to minimize a regularized least-squares criterion defined as
\begin{equation}
\label{eq:minimization}
\underset{x\in \mathcal{C}}{\operatorname{minimize}} \;\; \frac12 \| \mathcal{A}(x)-Y\|_2^2+\lambda R\big(\Psi^{\dagger} (x) \big).
\end{equation}
where  $R$ promotes sparsity of the target object in a basis induced by the operator $\Psi$ (e.g. wavelet basis \cite{Mallat_book}, gradient basis \cite{Rudin_1992_total_variation}, etc.). 
Moreover, the amplitude of $x$ is constrained to belong to $\mathcal{C}\subset\mathbb{R}^{N\times N}$. Finally, $\lambda>0$ is the regularization parameter balancing the importance of the regularization term  with respect to the data-fidelity term (least-squares criterion). \\
To compute the forward propagation $\mathcal{A}(x)$ efficiently, we separate the linear projection into two linear operations; computation of the spatio-temporal intensity field at the depth of the object and propagation of the light field from the object to the observation plane. The former can be performed by an element-wise multiplication between the object, or its current estimate, and each temporal frame of the light field propagated from the illumination point to the object. This only needs to be computed once using equation Eq.~\eqref{eq3}. The latter is computed by convolving the result of the previous operation with the point spread function given in equation Eq.~\eqref{eq3}. Performing these two operations is significantly more efficient than computing the large matrix multiplication representing the full forward propagation. In addition, in order to avoid fitting to noise at the edges of the recorded data and consequentially causing Fourier transform artefacts when performing the aforementioned convolution operations, the measurements fidelity term $\| \mathcal{A}(x)-Y\|_2^2$ is minimised only over pixels that experience a signal to noise ratio in the recorded data $Y$ over a certain threshold. This selection can be performed by multiplication with a selection mask $\mathcal{M}(Y)$. The operator $\mathcal{A}(x)$ is then computed as\\
\begin{equation}\label{op_as_conv}
\mathcal{A}(x) = \mathcal{M}(Y) \times \left[\Phi(\vec{r}_x,t;\vec{r}_0,t_0) \odot \left( \Phi(\vec{r}_out,t;\vec{r}_{x,0},t) \times x_T \right) \right],
\end{equation}
\\ Where $\odot$ indicates a convolution, $\vec{r}_x = (x,y,d_1)$ is the set of 2D spatial coordinates at the illumination-object distance $d_1$, $\vec{r}_0$ is the illumination position at the input surface, $\vec{r}_out = (x,y,d_1+d_2)$ is the set of 2D coordinates at the output surface, $\vec{r}_{x,0} = (0,0,d_1)$, $x_T \colon \mathbb{R}^{N \times N \times T}$ is constructed by repeating the 2D object $x$ in the third dimension $T$ times and $\mathcal{M}(Y)_{i,j,1:T}$ is one if $\sum_k^T Y_{i,j,k} \geq s$  and zero otherwise, with $s$ being a real positive constant.
\\ It is important to emphasize that both $R$ and $\Psi^{\dagger}$ can vary with different prior information about $x$. 
 We use the prior knowledge that the objects we wish to image are both piece-wise constant and sparse to then employ a regularisation term composed of two different penalty functions $\lambda R (\Psi^{\dagger} (x) ) = \lambda_1 R_1 (\Psi_1^{\dagger} (x) ) + \lambda_2 R_2 (\Psi_2^{\dagger} (x) )$. The first penalty function induces total-variation (TV) regularization by choosing $\Psi_1^{\dagger} (x) = [ D_h (x), \, D_v (x)]$, where $D_h  $ and $D_v $ represent the horizontal and vertical discrete gradients of the image, respectively \cite{Rudin_1992_total_variation}. 
  This regularization term is 
  given by
\begin{align}
\label{eq:tv}
R_1 \big(\Psi_1^{\dagger} (x) \big)
&	=	\| {x}\|_{\text{TV}} \nonumber\\
&	=	\sum_{i,j} \sqrt{ |x_{i+1,j}-x_{i,j}|^2 + |x_{i,j+1}-x_{i,j}|^2 }.
\end{align}
\\ The second penalty function induces sparsity by minimising the $\ell_1$ norm and is given by
\begin{align}
\label{eq:tv}
R_2 \big(\Psi_2^{\dagger} (x) \big)
=	\| {x}\|_{1} =	\sum_{i,j}  |x_{i,j}|.
\end{align}
\\ To solve the minimisation problem, we implement a steepest descent algorithm, which iteratively updates a solution $x^{(k)}$ with the gradient of the objective function in Eq.~\eqref{eq:minimization} as $x^{(k)} = x^{(k-1)} - aD^{(k-1)}$, where $D^{(k-1)}$ is the numerical gradient of the objective function evaluated at $x^{(k-1)}$ \cite{steep}.

%

\bibliography{biblio}
\bibliographystyle{naturemag}

\end{document}